\documentclass[graybox]{svmult}

\usepackage{mathptmx}       % selects Times Roman as basic font
\usepackage{helvet}         % selects Helvetica as sans-serif font
\usepackage{courier}        % selects Courier as typewriter font
\usepackage{type1cm}        % activate if the above 3 fonts are
\usepackage{makeidx}         % allows index generation
\usepackage{graphicx}        % standard LaTeX graphics tool
\usepackage{multicol}        % used for the two-column index
\usepackage[bottom]{footmisc}% places footnotes at page bottom

\makeindex             % used for the subject index
\begin{document}

\title*{Radio pulsar populations}
\author{Duncan R. Lorimer}
\institute{Department of Physics, West Virginia University, 
Morgantown, WV 26506, USA, \email{Duncan.Lorimer@mail.wvu.edu}}
\maketitle

\abstract{The goal of this article is to summarize the current state
of play in the field of radio pulsar statistics. Simply put, from the
observed sample of objects from a variety of surveys with different
telescopes, we wish to infer the properties of the underlying sample
and to connect these with other astrophysical populations (for example
supernova remnants or X-ray binaries). The main problem we need to
tackle is the fact that, like many areas of science, the observed
populations are often heavily biased by a variety of selection
effects. After a review of the main effects relevant to radio pulsars,
I discuss techniques to correct for them and summarize some of the
most recent results. Perhaps the main point I would like to make in
this article is that current models to describe the population are far
from complete and often suffer from strong covariances between input
parameters. That said, there are a number of very interesting
conclusions that can be made concerning the evolution of neutron stars
based on current data. While the focus of this review will be on the
population of isolated Galactic pulsars, I will also briefly comment
on millisecond and binary pulsars as well as the pulsar content of
globular clusters and the Magellanic Clouds.}

\section{Selection effects in radio pulsar surveys}\label{sec:selfx}

The current sample of radio pulsars is now close to 2000 and is
continuously increasing thanks to a wide variety of large-scale and
targeted searches being carried out at most of the major radio
observatories. The approximate rate of discoveries at the current time
is about 100 pulsars per calendar year, and we expect this trend to
continue and accelerate over the next decade as more powerful
facilities come online (both at radio and non-radio wavelengths).  An
excellent example of recent progress can be seen in the flurry of
radio pulsar counterparts to Fermi gamma-ray sources as reported by
Ray and Saz Parkinson elsewhere in these proceedings. It is important
to note, however, that while this sample represents a great
improvement over, say, the situation 20 years ago, it still likely
only amounts to a few percent of the underlying population of pulsars
whose properties we wish to constrain.  The main observational
selection effects that cause this are summarized below.

\subsection{Flux--distance relationship} Like all astronomical sources,
observed pulsars of a given luminosity $L$ are strongly selected by
their apparent flux density, $S$.  In a classical Euclidean model, for
a pulsar a distance $d$ from Earth which beams to a certain fraction
$f$ of $4\pi$ sr, the flux density $S = L/(4\pi d^2 f)$.  This is
known as the inverse square law and is commonly assumed for
astrophysical sources.  Since all pulsar surveys have some limiting
flux density, only those objects bright or close enough will be
detectable. Note that in the absence of prior knowledge about beaming,
geometrical factors are usually ignored and the resulting
`pseudoluminosity' is quoted at some standard observing frequency;
e.g., at 1400 MHz, $L_{1400} \equiv S_{1400} d^2$.  Recently, the
validity of the inverse square law has been called into question by
Singleton et al.~\cite{ssm+09}, 
and that perhaps the flux scales as $1/d$ instead.
It is important to fully investigate this claim. As Singleton et al.~point
out, if confirmed, it would have dramatic implications for many of
the conclusions presented here. In a search for radio transients
in M31 with Westerbork, Rubio-Herrera~\cite{rub10}
 has investigated the implications
of non-$1/d^2$ scalings on his results
and finds that many more transients should have
be observable for a $1/d$ law, and that a $1/d^2$ law is consistent with
the number of candidates seen. For now, we note
that any flux--distance relationship will bias the sample towards
bright and/or nearby objects.

\subsection{The radio sky background} A fundamental sensitivity limit
on any radio observation
is the system noise temperature, normally expressed
in Kelvins as $T_{\rm sys}$.
While every effort is made to minimize this at the telescope,
synchrotron radiating electrons in the
Galactic magnetic field contribute significantly with a `sky
background' component, $T_{\rm sky}$. At observing frequencies $\nu
\sim 0.4$ GHz, $T_{\rm sky}$ dominates $T_{\rm sys}$ for
observations along the
Galactic plane.  Fortunately, $T_{\rm sky} \propto \nu^{-2.8}$ so this
effect is significantly reduced when $\nu > 0.4$~GHz.

\begin{figure}[hbt]
\includegraphics[width=\textwidth]{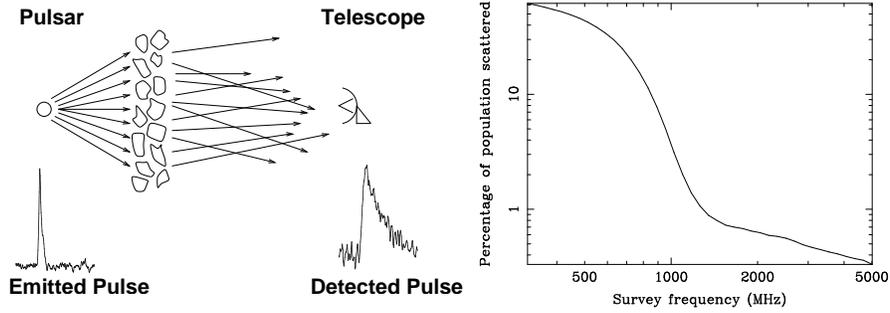}
\caption{\label{fig:scat}
Left: pulse scattering by irregularities in the interstellar medium
shown here as an idealized `thin screen' of material lying midway 
between the pulsar and the observer.
Right: a simulation showing the 
fraction of pulsars undetectable due to
scattering as a function of observing frequency.}
\end{figure}

\subsection{Propagation effects in the interstellar medium} Dispersion
and scatter-broadening of the pulses in the interstellar medium
hamper detection of
short period and/or distant objects. The effects of scattering are
shown in Fig.~\ref{fig:scat}.  Fortunately, like $T_{\rm sky}$, the
scatter-broadening time $\tau_{\rm scatt}$ has a strong frequency
dependence, scaling roughly as $\nu^{-4}$.
Fig.~\ref{fig:scat} shows that for survey frequencies below
1 GHz, scattering `hides' a large fraction of the
population. Additionally, scintillation, the diffractive
and refractive modulation of apparent flux densities by turbulences
in the interstellar medium \cite{ric70}
affect pulsar detection.
For example, two northern sky surveys carried out
20 years apart with comparable sensitivity \cite{dth78,snt97}
detected a number of pulsars above
and below the nominal search thresholds of one experiment but not the
other.  Surveying the sky multiple times minimizes the
effects of scintillation and enhances the detectability of 
intrinsically faint pulsars.

\subsection{Finite size of the emission beam} 
The fact that pulsars do not beam to $4\pi$ sr means that we see only
a fraction $f$ of the total active population. For a circular 
beam, Gunn \& Ostriker \cite{go70} estimated $f \sim 1/6$.  A consensus on
the precise shape of the emission beam has yet
to be reached. Narayan \& Vivekanand \cite{nv83} argued that the beams
are elongated in the meridional direction. Lyne \& Manchester \cite{lm88},
on the other hand, favour a circular beam. Using the same database,
Biggs \cite{big90} presented evidence in favour of meridional compression!
All these studies do agree that the beam size is period dependent,
with shorter period pulsars having larger beaming fractions. A very
popular model assumed by current studies derives from the work of 
Tauris \& Manchester \cite{tm98} who found that
$f \simeq 0.09\left[\log(P/{\rm s})-1\right]^2+0.03$, where $P$ is the
period.  A complete model for $f$ needs to account for
other factors, such as evolution of the inclination angle between the
spin and magnetic axes and the beaming of millisecond pulsars.

\subsection{Pulse nulling} The abrupt cessation of the pulsed
emission for many pulse periods, was first identified by Backer
\cite{bac70}. Ritchings \cite{rit76} subsequently
presented evidence that the incidence of
nulling became more frequent in older long-period pulsars, suggesting
that it signified the onset of the final stages of the neutron star's
life as an active radio pulsar. Since most pulsar surveys have short
($<$ few min) integration times, there is an obvious selection effect
against nulling objects. Means of reducing the impact of
this effect are to look
for individual pulses in search data \cite{nic99}, survey the
sky many times, or use longer integrations. Indeed, the longer
dwell times (35-minute pointings) used
in the Parkes multibeam survey have been particularly successful in
this regard, discovering a number of nulling pulsars \cite{wmj07}.

\subsection{Intermittency}

Recently, a new class of ``intermittent pulsars'' has been
found. These provide unique and new insights into neutron star physics
and populations \cite{klo+06}. The prototype, PSR~B1931+24, shows a
quasi-periodic on/off cycle in which the spin-down rate increases by
$\sim 50$\% when the pulsar is in its on state compared to the off
state! While the behaviour of this pulsar appears to be linked to the
increase in magnetospheric currents when it is on, there is no
satisfactory explanation for this effect.  Since PSR~B1931+24 is only
visible for 20\% of the time, we can readily estimate that there
should be at least five times as many similar objects. We believe this
number may be severely underestimated. It is important to establish
how many similar objects exist, and what the related timescales of
their non-emitting state are. These pulsars, and their cousins the
rotating radio transients (discussed in Section 4.1), remain a very
exciting area of current research.

\section{Correcting the biases in the observed sample}

How can we account for the effects discussed above and recover the
properties of the underlying pulsar populations? While some progress
can be made analytically (see, e.g. the early work of Gunn \& Ostriker 
\cite{go70}),
the non-uniform nature of all the above effects more readily lends itself
to a Monte Carlo approach to modeling pulsar populations and their
detection. The two main ways to implement such models can be thought
of as either a fully dynamical approach or a static ``snapshot'' model.
For the former case, a simulation is created in which a model galaxy
of pulsars is seeded according to various prescriptions of birth locations
and initial rotational parameters. Each of these synthetic pulsars is
then ``evolved'' both kinematically in a model for the Galactic gravitational 
potential and rotationally using a model for neutron star spin-down. The
properties of the resulting population are then saved.

\clearpage
\begin{figure}[ht]
\includegraphics[width=\textwidth]{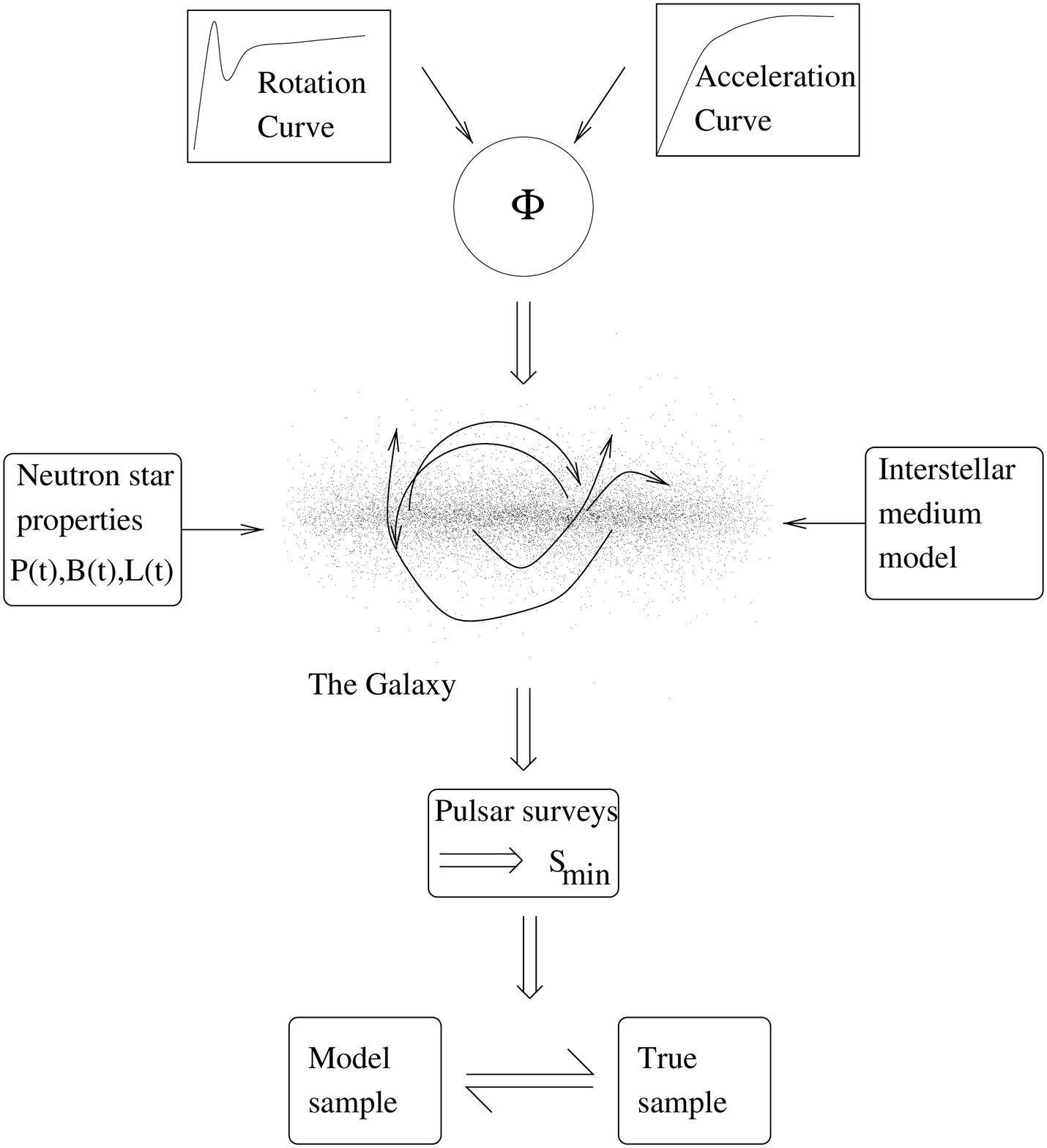}
\caption{\label{fig:monte} Schematic summarizing 
a fully dynamical Monte Carlo simulation of
the Galactic pulsar population. The main ingredients are
the model gravitational potential ($\Phi$), some prescription
for the neutron star evolution with time and a model of
the interstellar medium. Pulsars whose apparent flux densities
exceed those of the main surveys ($S_{\rm min}$) are saved
and the resulting ``model sample'' is compared to the 
actual sample of pulsars detected by those surveys.
}
\end{figure}
\clearpage

Using detailed
models for the pulsar surveys, it is possible to compute whether each synthetic
pulsar is actually detectable and the properties of these ``observable''
pulsars are saved. These samples may then be compared to the real observed
sample to asses the validity of the Monte Carlo model. The process is
summarized in Fig.~\ref{fig:monte}.
The snapshot approach
differs from the fully dynamical approach in that the pulsars are seeded 
at their final positions in the model galaxy without assuming anything
about spin-down or kinematic evolution and thus form a picture
of the current-day population. To model pulsar detectability, both
approaches are based around the well-known pulsar radiometer equation
\cite{dtws85} which has been demonstrated to provide 
an adequate description of the sensitivity of pulsar surveys 
\cite{kjv+10}.

The advantage of the snapshot approach over the dynamical one is that
it is simpler, requiring fewer assumptions about motion in the galaxy
or spindown and can often be optimized to form a model with a unique
best solution. Its major downfall, however, is that its simplicity means
that it says very little if anything about the progenitor population.
Fully dynamical models provide insights into these details, and can
for example describe the distribution of pulsars in $P$ and ${\dot P}$
space about which the snapshot approach is blind to. However, as discussed
below, a major point to keep in mind is that there is often no unique 
model that can describe the data and some care needs to be exercised
when interpreting the conclusions.

\section{Recent results}

With these caveats in mind, we now briefly review some of the latest
findings of studies which adopt either the snapshot or full dynamical
modeling approach.

\subsection{Pulsar space distribution}

\begin{figure}[hbt]
\includegraphics[width=\textwidth]{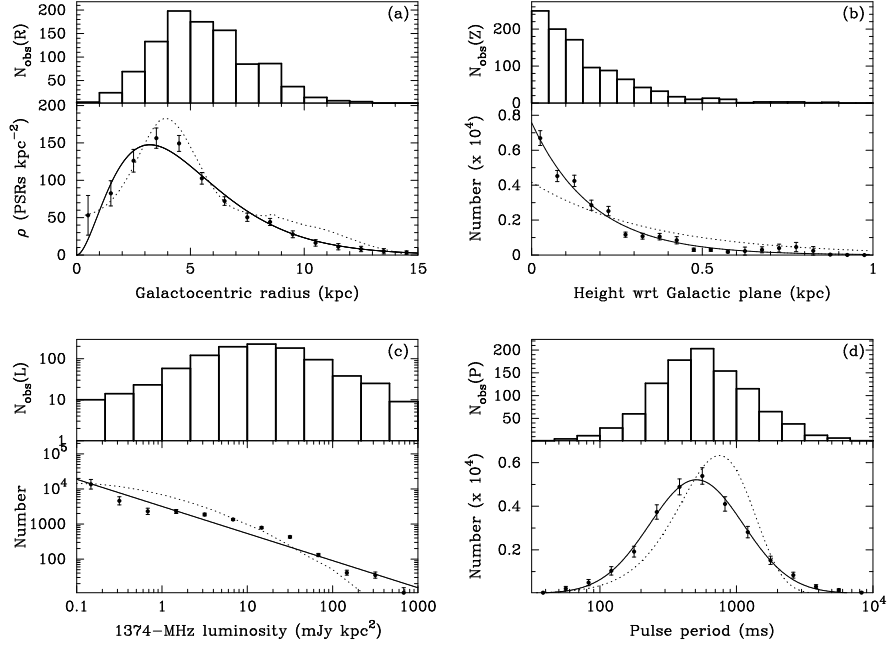}
\caption{\label{fig:pmps}
Observed number distribution from our input sample (upper
panels) and derived distributions (lower panels) for the
parameters: (a) $\rho(R)$; (b) $z$; (c) $L$; (d) $P$. The solid curves
are smooth analytic functions fitted to the data (see \cite{lfl+06}
for details).  The dotted curves show: (a) the assumed radial density
function of free electrons (from the NE2001 electron density model); 
(b) an exponential $z$ distribution
with a scale height of 350 pc; (c) a log-normal fit to the optimal
pulsar population model derived by \cite{fk06};
(d) a period distribution derived from studying pulse-width statistics~\cite{kgm04}.
}
\end{figure}

Models of the Galactic distribution of pulsars have been constructed
from observationally biased samples for many years \cite{dls77,tm77,lmt85}.
These studies typically follow the snapshot approach in which 
the population can be represented in terms of four independent distribution
functions: Galactocentric radius $R$, vertical dispersion from the Galactic
plane $z$, pulse period $P$ and luminosity $L$. In a recent approach
of this kind \cite{lfl+06}, we investigated models which accounted
for the observed distribution of pulsars seen by the Parkes Multibeam Pulsar
Survey \cite{mlc+01} which provides the largest uniform
sample (over 1000 Galactic pulsars) for such analyses. Using an
iterative Monte Carlo approach, we found that it is possible to find a 
unique model which converges to the same functional form regardless of the
initial shape of the distribution functions in $R$, $L$, $z$ and $P$. An
example of the model output is shown in Fig.~\ref{fig:pmps} which contrasts
the underlying and observed distribution functions for the final model.
The $L$, $z$ and $P$ distributions show the number of pulsars as a function
of each parameter. For $R$, the results are shown as the projected surface
density of objects on the Galactic plane, $\rho(R)$.

One important limitation of
this approach is that the form of the spatial distributions
$R$ and $z$ depends heavily upon the assumed model for the Galactic 
distribution of free electrons. The model shown in Fig.~\ref{fig:pmps}a
assumes the commonly used ``NE2001'' model \cite{cl02}. An
example of this dependence is the $R$ distribution in which the pulsars
naturally follow the $R$ distribution of free electrons. While the NE2001
model achieves a high level of sophistication, including electron
density enhancements in spiral arms, and can account for a wide variety
of observations, it is known to have a number of shortcomings 
\cite{kbm+03,gmcm08} which
are currently being addressed in a new model (Cordes, private communication).
In Fig~\ref{fig:pmps}b, for example, it is seen that the optimal model
$z$ distribution is significantly larger than observed --- this is a
direct result of the NE2001 electron scale height~\cite{lfl+06}.
It is conceivable that future population studies with larger samples of 
pulsars could be carried out where the distribution of free electrons
is allowed to vary. At the current time, however, one should be mindful of
the fact that any conclusions about the spatial distribution of pulsars 
are strongly coupled to models of the free electron density.

\subsection{Pulsar velocities}

A number of studies of the birth velocities of pulsars have been
carried out over the years and there has been much debate as to 
whether the distribution for nonrecycled pulsars
is unimodal \cite{ll94} or bimodal \cite{no90,acc02}
Recent studies \cite{hllk05,fk06} find no compelling evidence to
model the distribution with multiple components and the individual
1-D components of the pulsar's birth velocity vector follow either
a Gaussian \cite{hllk05} or exponential form 
\cite{fk06} with a mean value in the range 400--500~km~s$^{-1}$.

\begin{figure}[ht]
\includegraphics[width=\textwidth]{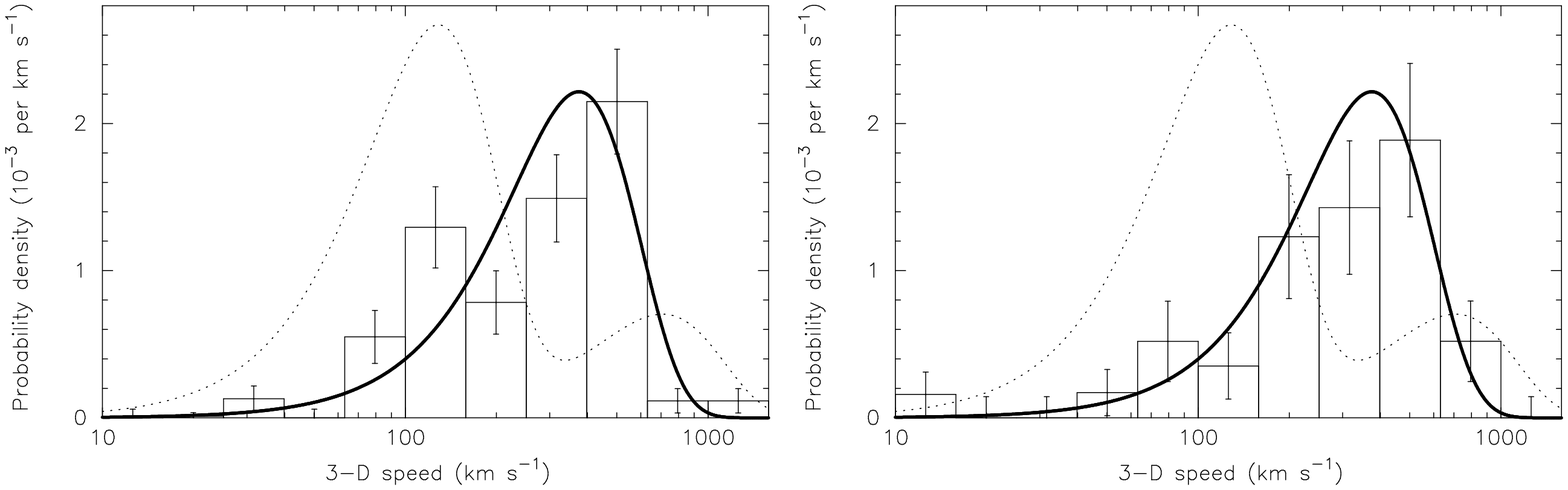}
\caption{\label{fig:v3d}
Normalized 3-D velocity probability density functions obtained from the
observed 1-D (left) 2-D (right) distributions using a
deconvolution technique \cite{hllk05}.  The uncertainties
on each histogram bin are calculated as the square root of
the number of pulsars in each bin. The dotted curve shows the 3-D
distribution favoured by Arzoumanian et al.~\cite{acc02}. The solid
curve is the best-fitting Maxwellian distribution
to the histogram from the 2-D distribution with
$\sigma=265$ km~s$^{-1}$.
}
\end{figure}

Fig.~\ref{fig:v3d} shows the results of a deconvolution
process from Hobbs et al.~\cite{hllk05} where the 3-D space
velocity distribution may be derived self consistently by
appropriately deprojecting
either the 1-D or 2-D distributions of young pulsars
(defined to be those with characteristic ages less than 1 Myr).
As can be seen, a previously suggested two-component model
is not implied by these data. While it is found for millisecond
and binary pulsars that the velocity distribution is different
to the normal pulsars shown here, the main conclusion to take
away from current results is that the distribution of velocities
for isolated radio pulsars is unimodal.

\subsection{Pulsar luminosities}

Because of the strong connection between distance and luminosity,
any uncertainty in the pulsar distance scale propagates through
to an uncertainty in the luminosity function \cite{mor81}. Two
critical questions concerning pulsar luminosities we wish to 
answer are: (1) what, if any, evolution in luminosity is there with pulsar
age? (2) what is the shape of the luminosity function?
The idea of a decay in luminosity has been in the literature for
some time. Taylor \& Manchester \cite{tm77} have pointed
out that the simple fact that the distribution of pulse periods
tails off at long periods demands that the luminosity decays with
time. If the luminosity were constant, many more pulsars would be
observed. This can be readily shown via simulations in which assigning
the luminosity to a pulsar at random results in a pile up of 
pulsars at high $P$ and low $\dot{P}$ which is not observed in the
real sample \cite{fk06}.

The exact form of the luminosity decay remains contentious, however.
While the best dynamical models can account for the observed data
with a simple power law model in which $L \propto P^{\alpha} 
\dot{P}^{\beta}$, the values of the exponents $\alpha$ and 
$\beta$ are neither readily found from fits to the observed population
\cite{lbdh93} nor are uniquely constrained from
the dynamical modeling \cite{rl10a}.

\begin{figure}[ht]
\includegraphics[width=\textwidth]{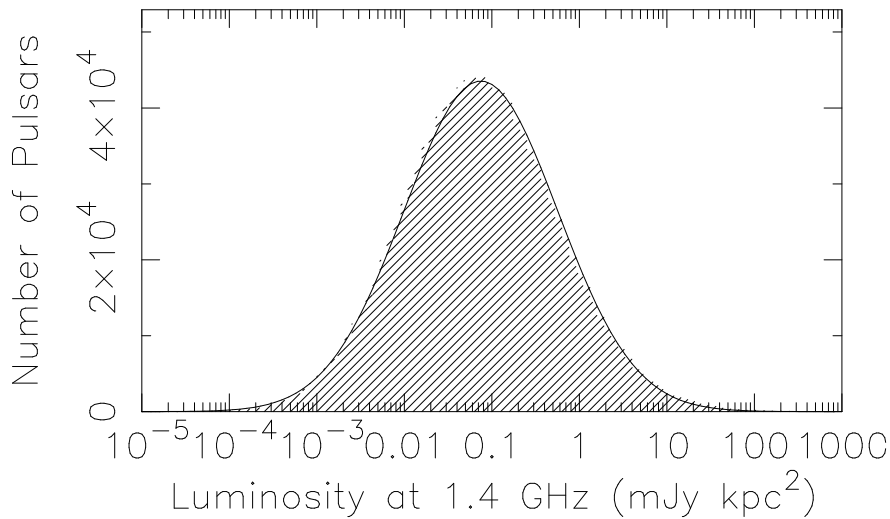}
\caption{\label{fig:lnorm}
Results from Faucher-Gigu\'ere \& Kaspi \cite{fk06} which
show the underlying luminosity function (defined to be at a
frequency of 1.4~GHz) for their optimal model
of the isolated pulsar population. The solid line shows a
Gaussian fit to the data where  the mean of the distribution
in log $L$ is --1.1 and the standard deviation is 0.9.}
\end{figure}

One result that does appear to be robust is the form of the luminosity
distribution. While the snapshot models typically favour some sort
of power-law distribution for the number of pulsars $N(L)$ in which
$d \log N / d \log L \sim -1$, they remain agnostic about the distribution
of luminosities {\it below} the minimum value in the observed sample,
$L_{\rm min}$. The dynamical approach suggests that the underlying
shape of the luminosity function is log-normal in form \cite{fk06}.
The parent
luminosity distribution from this simulation is shown in Fig.~\ref{fig:lnorm}.
Simulations with different spin-down
models all appear to show the same basic shape \cite{rl10a,ppm+10}
Whether this distribution applies to millisecond
pulsars is currently unclear.

\subsection{Magnetic alignment}

\begin{figure}[ht]
\includegraphics[width=\textwidth]{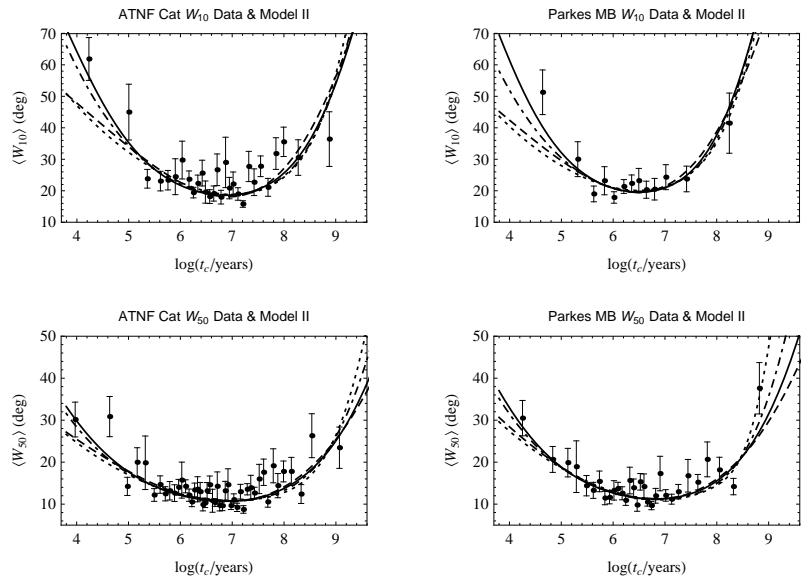}
\caption{\label{fig:align}
Results of model fits by Young et al.~\cite{ycb+10} to various pulse-width
characteristic age relations for Parkes multibeam data. All the models
shown include the effects of magnetic alignment on a timescale of a few
million years. The increase of pulse widths at large characteristic ages
appears only to be explained by the alignment process.
}
\end{figure}

Recently, two groups have provided strong evidence that the angles between
the magnetic and spin axes of neutron stars are not random and, in fact,
appear to decay on a timescale of $10^7$~yr or less. Weltevrede \& Johnston
\cite{wj08} provide strong empirical evidence for such magnetic alignment
based on the statistics of pulsars which exhibit interpulses. They point
out that the fraction of pulsars whose profiles can be described by 
viewing an orthogonal rotator is strongly linked to the stars' rotational
period, with a much higher interpulse fraction observed at longer periods
than would be expected from randomly inclined lighthouse beams. Unless
the observational sample is in some way biased, their conclusions appear
to be irrefutable.
In an independent approach, Young et al.~\cite{ycb+10} also
find evidence for magnetic alignment from an analysis of the pulse
width statistics of pulsars. They argue, from graphs of pulse width
versus characteristic age (see examples in Fig.~\ref{fig:align})
which show a turn-up at long periods, that only alignment
on a timescale of a few million years can explain the increase in pulse
width. The two competing effects which shape these curves are the
narrowing of pulse widths with period, and the alignment of the magnetic
axis which means that older pulsars are more likely to be seen
as aligned objects where the emission occupies a larger fraction
of the rotational period.

Both the above studies suggest that some sort of alignment is taking place
in isolated radio pulsars. As pointed out by Ridley \& Lorimer \cite{rl10a}
however, this observation throws up a conundrum when one
attempts to construct a self-consistent model of spin-down evolution.
The standard magnetic dipole model, in which the braking torque
is proportional to the square of the sine of the inclination angle,
does not do a good job of reproducing the $P-\dot{P}$ diagram if
this angle evolves with time in the manner expected above. Furthermore,
the hybrid spin-down model of Contopoulos \& Spitkovsky \cite{cs06} which 
can account for alignment, appears to provide a very poor description of the
observed distribution in $P-\dot{P}$ space. A more sophisticated
model for pulsar spindown is needed which can reconcile these differences.

\subsection{Magnetic field decay}

No discussion of pulsar statistics would be complete without a mention
of magnetic field decay --- a contentious issue that
has raged for the past 30~yr. For many years, it was believed that 
the magnetic fields of isolated pulsars decayed exponentially on a $1/e$
timescale of 10~Myr or less \cite{lmt85}. Popular
opinion switched toward favouring models with essentially no field decay
in the 1990s \cite{bwhv92,lbh97}.

\begin{figure}[ht]
\includegraphics[width=\textwidth]{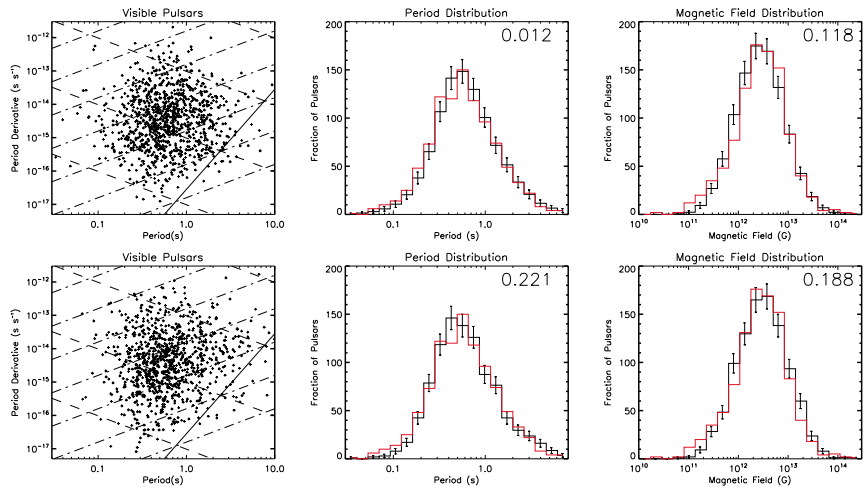}
\caption{\label{fig:decay}
Results of simulations by Popov et al.~\cite{ppm+10} which show 
how comparable results can be found by modeling the population
without field decay (upper panels, following the prescription
given by Faucher-Gigu\'ere \& Kaspi \cite{fk06}) and with field
decay (lower panels, following a magneto-thermal model of
Popov et al.~\cite{ppm+10}). Both models give statistically equivalent results.
The observed distributions in period and magnetic field strength
are shown by the red histograms. Kolmogorov-Smirnov probabilities 
that the model and observed data are drawn from the same parent population
are shown on the top right hand corner of each histogram.
The model distributions are
shown with statistical error bars. Both models appear to provide equally
viable descriptions of the data.
}
\end{figure}

Nowadays, the community is split between no significant magnetic field
decay \cite{fk06,rl10a} and 
a decaying field \cite{gob+02,ppm+10}. The simulations shown
in Fig.~\ref{fig:decay} present models of the radio pulsar population
with and without the effects of magnetic field decay. Is this
another example of covariance between model parameters, or fundamental
differences in modeling techniques? For me, the answer remains to be found.

\section{Final thoughts and future prospects}

In this review, I have focused on a number of recent results concerning
the population of isolated pulsars. While we have come a long way in 
understanding the distribution of these object in the Galaxy, their
initial velocity dispersion and luminosity function, much remains to be
understood in terms of their spin-down behaviour. A model which can
account for the observed magnetic alignment, nulling, beaming
and spin-down evolution 
is a major goal for any future study. Some further areas that are ripe
for research are summarized briefly below.

\subsection{Rotating radio transients}

An even more extreme class of intermittent neutron stars are the
so-called rotating radio transients \cite{mll+06}.  Their detection
was made possible by searching for dispersed radio bursts \cite{cm03}
which often do not show up in conventional Fourier-transform based
searches \cite{lk05}.  Since the initial discovery, a significant
effort has gone in to searching for and characterizing more RRATs.
Over 30 are currently known \cite{hrk+08,dcm+09,kle+10,bb10,mlk+09} but only
seven have timing solutions, with four of these only recently achieved
\cite{mlk+09}.  Recently, Lyne et al.~\cite{lmk+09} reported the
detection of two glitches in RRAT J1819$-$1458. While these events are
similar in magnitude to the glitches seen in young pulsars and
magnetars, they are accompanied by a long-term {\it decrease} in the
spin-down rate, suggesting that it previously occupied the phase space
populated by the magnetars.  Further observations are needed to
confirm this ``exhausted magnetar'' hypothesis. The Galactic population of such
objects is potentially significant and it remains to be determined
whether alternative evolutionary scenarios need to be invoked other
than core-collapse supernova \cite{kk08}. Further work will certainly
clarify this issue as known sources are better characterized.

\subsection{Millisecond pulsars}

For many years, studies of the Galactic population of millisecond
pulsars have been plagued by small-number statistics \cite{kn88}.
More meaningful results were obtained during the 1990s
with the advent of all-sky surveys of the local population \cite{lml+98}
where it was found that the velocity distribution of
millisecond and binary pulsars is significantly lower than that of
the normal isolated population discussed above. Presently, with
an exponentially growing sample of millisecond pulsars we are in
an era where it is possible for the first time to carry out full
population syntheses of the Galactic population. Story et al.~\cite{sgh07}
have carried out much work in this area and have paved the way for future
studies, though many questions 
remain to be answered including: (i) what is the overall Galactic
distribution of millisecond pulsars?; (ii) is the millisecond pulsar
luminosity function comparable to normal pulsars?; (iii) are all
millisecond pulsars produced in low-mass X-ray binary systems?;
and (iv) what is the origin of isolated millisecond pulsars?

\subsection{Pulsars in the Magellanic Clouds}

Currently 19 radio pulsars are known in the Large and Small
Magellanic Clouds \cite{mfl+06}. Ridley \& Lorimer \cite{rl10b}
recently carried out a snapshot analysis of this population assuming
the log-normal luminosity function for Galactic pulsars 
described above. We found that there are roughly 18,000 and  11,000
normal pulsars in the large and small clouds respectively.  After accounting
for beaming effects, and the fraction of high-velocity pulsars
which escape the clouds, the estimated birth rates in both clouds
appear to be comparable and in the range 0.5--1 pulsar per century.
Although higher than estimates for the rate of core-collapse
supernovae in the clouds, these pulsar birth rates are consistent with
historical supernova observations in the past 300 yr.  A fully
dynamical model incorporating the kinematics and spindown of the 
pulsars in the Magellanic Clouds would be a logical extension of this work.

A substantial population of active radio pulsars (of order a few
hundred thousand) have escaped the clouds and populate the 
local intergalactic medium. For the millisecond pulsar
population, the lack of any detections from current surveys leads 
only upper limits of up to 40,000 sources in the two clouds. 
A new survey with greatly improved time and frequency resolution
currently underway at Parkes could detect a few of these sources
(if they exist) and place valuable constraints on the total population.
Giant-pulse emitting neutron stars could also be seen by this survey.

\subsection{Globular cluster pulsars}

The first pulsar in a GC was found over 20 years ago~\cite{lbm+87}.
Currently, there are
140 radio pulsars in 26 GCs~\cite{gcpsrs}. Progress towards the
current sample has proceeded in two phases. In the late 1980s
and early 1990s, searches uncovered about two dozen of the brightest
objects~\cite{ka96}. 
Further progress was only made later, around the year 2000,
when a combination of advances in
high-frequency broadband receivers, software algorithms, computing
power and data storage capabilities led to a resurgence of
discoveries~\cite{clf+00,dlm+01,rgh+01,pdm+03,rhs+05} and
interest from observers and theorists~\cite{cr05}.

While much of the recent focus on the observational results mentioned
above has been on revealing unique systems and their applications for
fundamental physics, relatively little attention has been paid on
understanding the population of GC pulsars as a whole. In fact, the
last major study into GC pulsar statistics were carried out in the
late 1980s~\cite{knr90}. A major finding of this work was that the
birth rate required to sustain the population of $\sim 10^4$ MSPs
estimated in all GCs was 100 times higher than the birth rate of their
proposed progenitors~\cite{acrs82}, the LMXBs.  
Since then, the discovery of large numbers of quiescent
LMXBs\cite{hgl+03} has decreased this disparity, but another potential
problem has emerged.  If NSs are formed as in the Galaxy, i.e.~in the
core collapse supernovae of massive stars, then the large resultant
velocities observed among the young pulsars~\cite{hllk05} would
eject the vast majority of all NSs from GCs. This
would result in a very small number of primordial NSs in clusters.
How do GCs retain enough NSs to form all the quiescent LMXBs and MSPs
we observe? Are there other NS formation mechanisms at work?
We anticipate significant progress in many of these areas in the near
future.

\section*{Acknowledgments}

My research is funded by the West Virginia Experimental Program
to Stimulate Competitive Research, the Research Corporation, the
Smithsonian Astrophysical Observatory, the
National Radio Astronomy Observatory and the National Science
Foundation. I thank the Physics Department, Eberly College of Arts \&
Sciences and the Senate Research Committee for supporting my travel to
this meeting.

\end{document}